\documentclass[conference]{IEEEtran}
\IEEEoverridecommandlockouts
\usepackage{graphicx}
\usepackage{tabularx}
\usepackage[justification=centering]{caption}
\usepackage{latexsym}
\usepackage{amsmath}
\usepackage{algorithm}
\usepackage[noend]{algpseudocode}
\usepackage{booktabs}
\usepackage{multirow}
\usepackage{siunitx}
\usepackage{pdfpages}
\usepackage{booktabs}
\usepackage{afterpage}
\usepackage{float}
\usepackage{subfigure}
\usepackage{amsfonts}
\usepackage{amssymb}
\usepackage{cite}
\DeclareMathOperator{\supinf}{supinf}

\ifCLASSINFOpdf

\else
\fi

\begin{document}
%
\title{A Gossip Algorithm based Clock Synchronization Scheme for Smart Grid Applications}

\author{\IEEEauthorblockN{Imtiaz Parvez${^{*}}$, Arif I. Sarwat${^{*}}$, Jonathan Pinto${^{*}}$, Zakaria Parvez ${^{\dag}}$ and Mohammad Aqib Khandaker ${^{\ddag}}$ }
\IEEEauthorblockA{{${^{*}}$Department of Electrical and Computer Engineering, Florida International University, Miami, FL 33174.}\\ {${^{\dag}}$Electricity Generation Company of Bangladesh.}\\ {${^{\ddag}}$Department of Electrical and Electronic Engineering, Independent University, Dhaka, Bangladesh.}\\
Email: {\tt \{iparv001,asarwat,jpint011\}@fiu.edu,zakaria.parvez@egcb.com.bd,mak127@live.com}}%
 
\thanks{This research was supported in part by the U.S. National Science Foundation under the grant RIPS-1441223 and CAREER-0952977.}
}


\maketitle

\begin{abstract}
The uprising  interest in  multi-agent based networked system, and  the numerous number of applications in the distributed control of the smart grid leads us to address the problem  of time synchronization in the smart grid. Utility companies look for new packet based time synchronization solutions with Global Positioning System (GPS) level accuracies beyond  traditional packet methods such as Network Time Protocol (NTP). However GPS based solutions  have poor reception in indoor environments and dense urban canyons as well as GPS antenna installation might be costly. Some smart grid nodes such as Phasor Measurement Units (PMUs), fault detection, Wide Area Measurement Systems (WAMS) etc., requires synchronous accuracy as low as 1~ms. On the other hand, 1~sec accuracy is acceptable in management information domain. Acknowledging this, in this study, we introduce gossip algorithm based  clock synchronization method among network entities from the decision control and communication point of view. Our method synchronizes clock within dense network with a bandwidth limited environment. Our technique has been tested in different kinds of network topologies- complete, star and random geometric network and demonstrated satisfactory performance.

\end{abstract}

\vspace{2mm}

%
\IEEEpeerreviewmaketitle

\section{Introduction}

The smart grid is an innovative improvement of the conventional one way electric power system  to two directional power flow system utilizing the ICT communication and distribution systems to deliver electricity to consumers \cite{7794179, keymanPES}. The grid includes distribution side (smart meters network), transmission side (transmission lines, substation etc.), generation side (generators and other renewable sources), intergeneration of renew energies, and various technologies to integrate them  with improved monitoring, control, and efficiency. Bidirectional communication is used for  the interaction among  the entities of the smart grid \cite{p1,p2,p5}.  The control and monitoring of smart grid monitors and controls  the grid with outage management, demand response, disaster prevention-recovery and security \cite{5540266, psec6,parvez2015reliability,locationbased}. While control and monitoring the grid, synchronization is needed to maintain phase and frequency equal in the entire grid.


Clock synchronization is employed in the diverse sections of the smart grid for maintaining equal values of voltage and current. At present, Phasor Measurement Units (PMUs) provide accurate information about voltage, current, phase, frequency, and rate-of-change-of-frequency using a highly accurate timestamp \cite{7156067,7511666}. Gathering PMU data provides a definite and complete view of the entire grid. PMU measurements are taken around 30 observations per second, timestamping each measurement to a common reference time  from different locations and utilities. The time of those locations and utilities need to be synchronized. However, the PMUs need to be co-located to avoid communication delays. Wide Area Measurement Systems (WAMS)  are being employed in the power industry  for real-time measurements of different grid parameter \cite{7420696,roleofclockWAMS}. Current systems, such as SCADA (supervisory control and data acquisition)/EMS (energy management systems), do not provide efficient solutions in the case of cascading because these systems are designed to act locally. WAMS also employs PMUs in strategic locations through a wide area which allows for accurate monitoring in the case of a cascaded outage.

Traveling-wave fault detection is another method used by power companies to detect a fault in a line by measuring voltage and current surges that may be caused by disturbances such as  line fault, switching operations, and lightning \cite{travelwave}. Time elapsed is measured between fault position to measurement point. Protective relay schemes are also employed by engineers to protect transmission line. Time synchronization is vital in this implementation because of the need of accurate readings of 1 ms or less. Time based  billing also requires time synchronization in the order of milliseconds. Incentives are offered to customers for using power at off-peak hours by offering discounted prices. High power quality is expected from consumer utilities. However, certain loads such as harmonics and flicker can affect power quality. Incentives are offered for maintaining power quality. Accuracy of 1 ms is required to synchronize periodic measurements of power-quality quantities such as harmonics and flicker. Substation and power network calls for the merging of all Intelligent Electronic Devices (IEDs) to share data and control commands that enable support of efficient protection, monitoring, automation, metering, and control functions. IEC 61850  is the international standard for substation automation systems \cite{adamiak2010iec}. Time synchronization is vital in the IEC 61850-9-2 process bus and relies on accuracy of sub microseconds. On the other hand, information management  domain of the smart grid allows latency upto 1ms \cite{6307240}.

Global positioning system (GPS) is the most popular solution for the time synchronization problem which provides sub 100 nanosecond accuracies, and are often used where precision time and frequency synchronization is critical. However, improved accuracy comes at a cost. GPS based systems require outdoor antenna installations with direct view of the sky, which  not only adds extra expense but also increases an extra burden on the physical infrastructure of the facility. GPS has poor reception in indoor environments and dense urban canyons where direct visibility of the GPS satellites is poor. 

Packet-based synchronization solutions are increasing in demand as an alternative to GPS based solutions due to packet networks (Ethernet and IP) being used as the common mode of communication in power systems \cite{packetbased,packet2,packet3,p3,p4}. Network time protocol (NTP) \cite{MillsNTP} is a protocol designed to synchronize computer clocks over a packet network. Typical accuracy for the NTP server is one millisecond for  LAN (local-area network), and a few tens of milliseconds for WAN (wide-area network). However, compared to UTC (Coordinated Universal Time), accuracy will vary depending upon network traffic.

In this study, we introduce gossip algorithm \cite{NET014,Iwanicki2006,parvezaverage} based time synchronization for smart grid nodes such as PMU and smart meters from the decision control and communication point of view. Currently smart meter clocks are not synchronized. The out of step smart meters pose a challenge for utilities to leverage smart meter data for advanced distribution network operation and control, such as outage management \cite{7352376}. Since, smart meters are close located, our approach can be potential solution for clock synchronization. Consensus building by gossip algorithm has been analyzed for various multi agent networks such as sensor fusion, random networks, synchronization of coupled oscillators, algebraic connectivity of complex networks, asynchronous distributed algorithms, formation control for multirobot systems, optimization-based cooperative control, dynamic graphs, complexity of coordinated tasks and consensus based belief propagation in Bayesian networks etc.\cite{Iwanicki2006, 1498447,Angius}. In a bandwidth limited environment with numerous nodes, our technique provides a potential solution for clock synchronization. The technique makes consensus in time values gossiping randomly with neighbors invalidating the need of communication with all nodes simultaneously. Since the data packets for node synchronization are small, we propose pulse code modulation for transferring the data among nodes. We tested the performance of our proposed technique on difference types of network topologies such as complete, ring, and random geometric graphs. 

The rest of this paper is organized as follows: Section~II presents gossip algorithm based system model. In section~III, convergence time has been analyzed. Section~IV describes the simulation results. Finally, a brief conclusion is included in Section~V.

\section{Sytem model}

\begin{figure}[t!]
	\centering
	\includegraphics[width=1\linewidth]{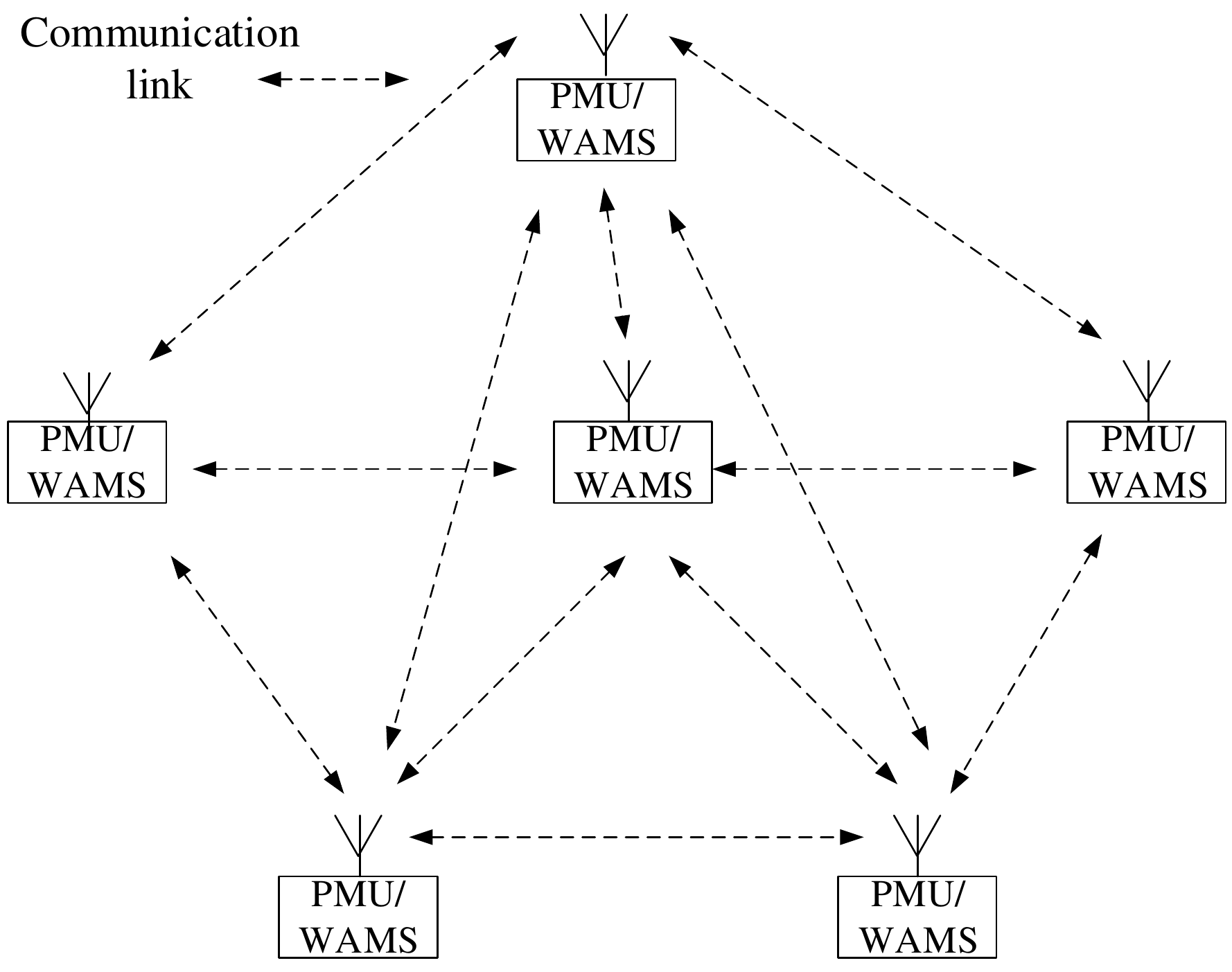}
	\caption{An example network of PMU or WAMS of smart grid.}
	\label{fig:roc}
\end{figure}

Let us consider, a smart grid network of $N$ nodes specified by graph $G$ (as illustrated in Fig. 1). The time of the network will be synchronized. For a undirected graph $G$, we define
\begin{equation}
G =(\upsilon\cal,\varepsilon\cal)
\end{equation}
where $\upsilon=\{1,2,....,N\}$ is the set of vertices (nodes), and \\
$\varepsilon\cal$ is the set of edges which is subset of $\{\{i,j\}: \{i,j\}\in\upsilon, i\neq j\}$.

Each node has finite value $t(l)\in\mathbb{R}$ where $l$ is the time instant/iteration number. Any two nodes $i\in\upsilon$ and $j\in\upsilon$ are connected through wireless quantized communication.
The sequences of paths of vertices $\upsilon\in G$ is $(i=i_1,i_2,......i_r=j)$ such that  
$\{i_j,i_{j+1}\}\in G $. The graph is fully connected or complete if $ \{(i,j):(i,j) \in \upsilon, i\neq j\}$. Since $G$ is a undirected graph, an edge $ (i,j) $ is selected randomly with equal probability $P$ such that $\sum\limits_{(i,j)\in \varepsilon\cal}  P_{(i,j)}=1$.

Let $P\in\mathbb{P}^{{N\times N}} $ be , then

\begin{equation}
P_{ij}= P_{ji} = \begin{cases}  P_{(i,j)} & \quad\text{if}\hspace{2mm}(i,j)\in \varepsilon\cal \\ 0 & \quad\text{otherwise}\end{cases} 
\end{equation}

The nodes are connected by wireless link. Since the graph is connected, there is always a path between any node $i$ and $j$. Therefore, a path in $G$ consists of a sequence of vertices $(i=i_1,i_2,......i_r=j)$ such that $\{i_j,i_{j+1}\}\in G $ for every $ j\in\{1,2.....r-1\}$. Since the nodes are connected by digital wireless link, they cannot extract the real value but the quantized value of exact value.

\begin{figure*} [ht!]
 	\centering
 	\subfigure[] { \label{fig:a}\includegraphics[width= 0.3\linewidth,height=5.5 cm]{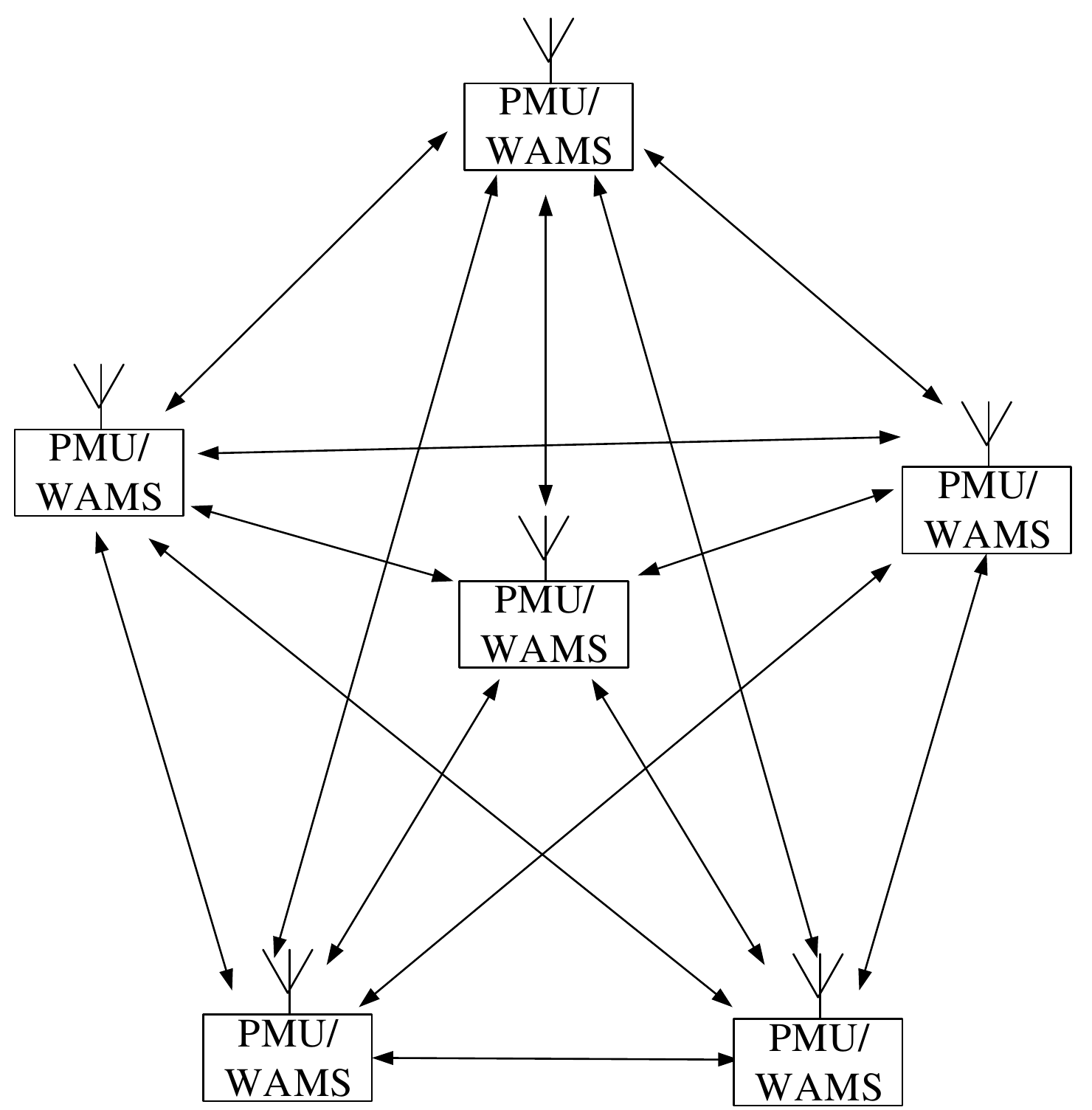}}
 	\subfigure[] {\label{fig:b}\includegraphics[width= 0.3\linewidth,height=5.5 cm]{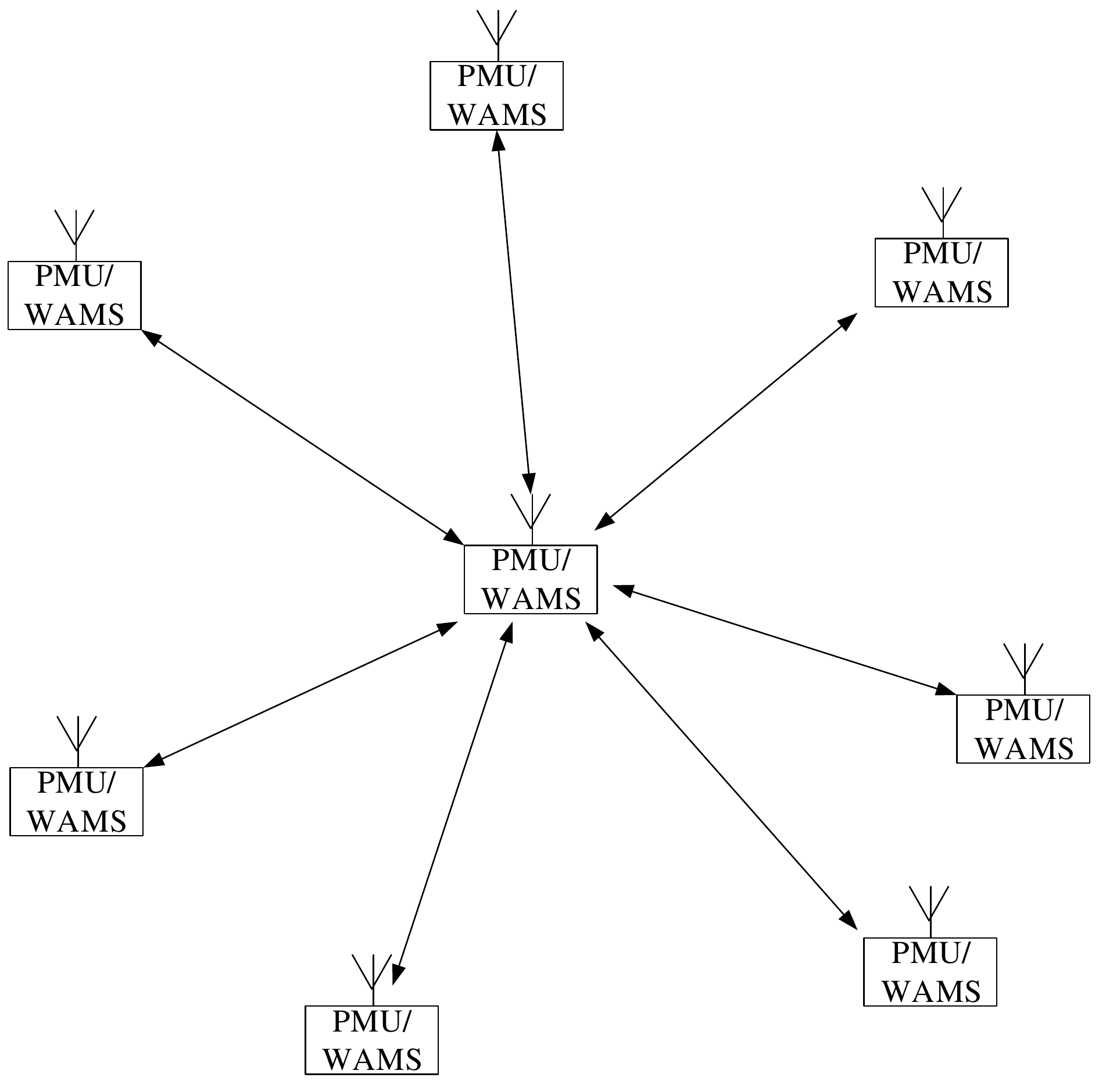}}
 	\subfigure[] {\label{fig:b}\includegraphics[width= 0.3\linewidth,height=5.5 cm]{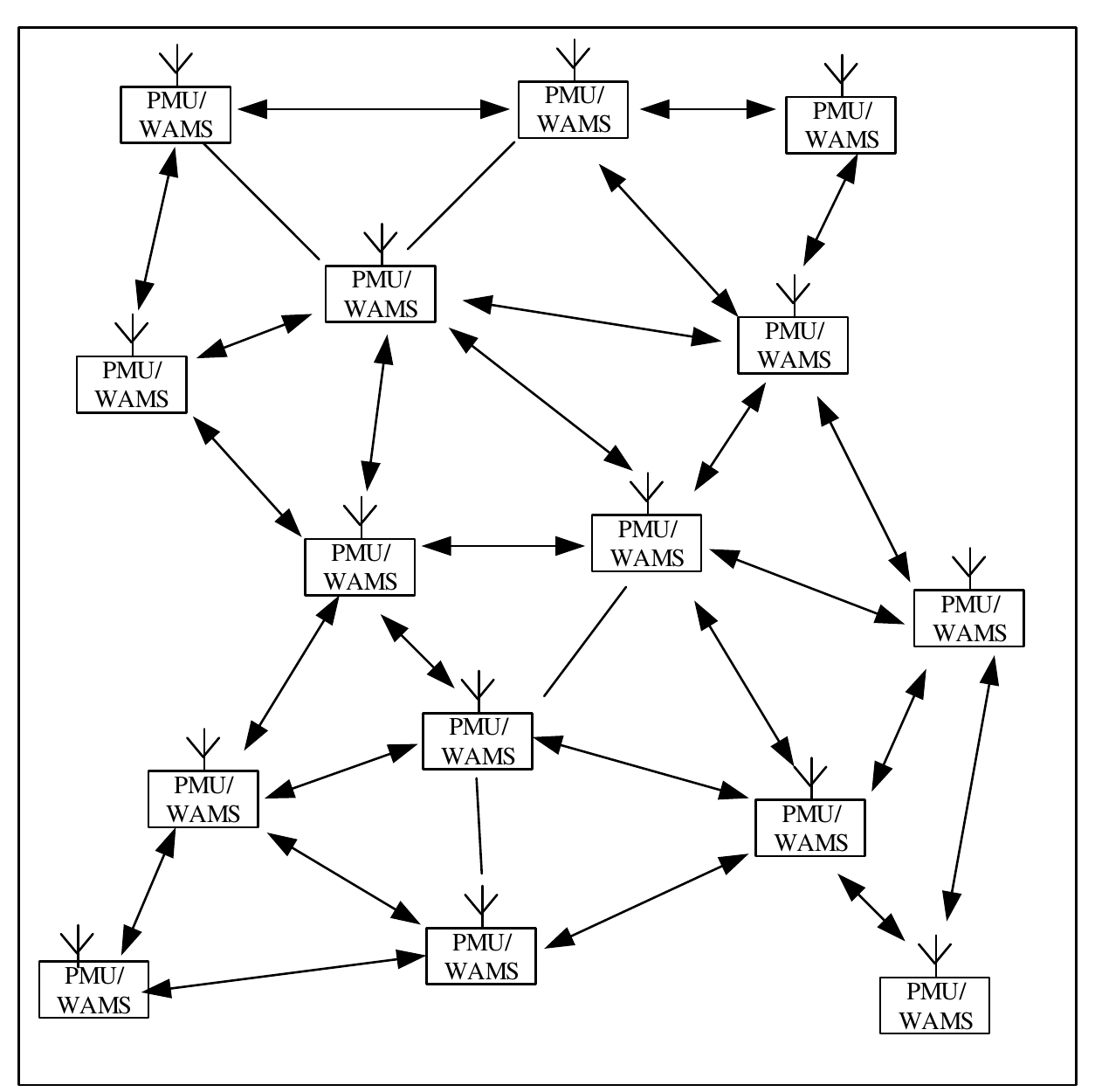}}	
 	\caption{Three kinds of network topologies (a) Complete graph network (b) Ring graph netwok (c) Random geometric graph network}
 	\label{fig:NE1}
 	\vspace*{-2mm}
 \end{figure*}

In our algorithm, the previous state value is preserved. On each time instant/iteration step $l$, two randomly chosen nodes update their value by using the below formula-

\begin{align}
t_i (l+1)= t_i(l)- 1/2Q[t_i(l)] + 1/2 Q[t_j(l)]
\\t_j (l+1)=t_j(l)-1/2 Q[t_j(l)]+1/2 Q[t_i (l)]
\end{align}

where $ Q[t(l)] $ is the quantized value of $t(l)$.

A network $G$ with initial node value $T(0)=\big(t_i(0)\big)_{i=1}^N$ where $T\in \mathbb{R}$ is reached to quantized consensus after $l$ iteration steps/time instant if and only if

\begin{equation}
\ | \bar{t_i}(l) -{N^{-1}} \sum\limits_{j=1}^{N} t_j (0)|< 1
\end{equation}

and

\begin{equation}
l>T_{con}
\end{equation}

where $T_{con}$ is the convergence time/iteration steps.

In the consensus state, every node has almost the same value $t_i(l) \simeq t_j(l) \simeq \bar{t} $$ \hspace{2mm}  \forall (i,j)\in G$. \\

\textbf{Proof of Convergence:} Let us consider that each node has value $t_i(l)\in \mathbb{R},$ $i\in \upsilon$ at time instant $l$. Further we define

$m (l)=\min\limits_{i} t_{i}(l)$ 

$M (l)=\min\limits_{i} t_{i} (l)$
 
$D(l)=|M(l)-m(l)|$

Based on the above definition, for $D(l)\ge 1$  we set the below updating rules:

(Rule 1) $t_i (l+1) + t_j (l+1) = t_i(l) +t_j(l) $

(Rule 2) If $ D_{ij} (l) > 1 $ then $D_{ij} (l+1) < D_{ij} (l) $

(Rule 3) If $ D_{ij} (l) = 1 $ and $t_i (l) > t_j (l)$, then $t_i (l+1)=t_j(l) $ and $ t_j (l+1)=t_i (l)$. This is called swap rule.

Since  the summation of  node values are constant i.e. $\sum\limits_{j=1}^ {N} t_j(0)=\sum\limits_{j=1 }^ {N} t(t)$, we assume that:

(A1) During the execution of gossip algorithm, node value lies in the finite set $\mathcal{J}$ for any time/iteration step $l$ given that initial node value is $t(0)$.

(A2) For any node value state $t(l)=t$, then there exists finite time instant or iteration step $k_l$ such that $Pr[t(l+k_t)\in S|t(l)] >0$ where $S$ is the set of all vectors containing quantized consensus distribution.

(A3) If $t(l)\in S$, then  $t(l')\in S$ for $l'>l$.

Since $\mathcal{J}$ is finite and from (A2), let 

\begin{equation}
I=\min Pr [t(l+k)\in S/t(l)=t]
\end{equation}

\begin{equation}
K=\max_l k_l
\end{equation}

From property of [A3], its follows that 

\begin{equation}
Pr[t(l+k)\notin S | t(l)\notin S]\le(1-I)
\end{equation}

So, for $l$ number of time instant/iteration steps

\begin{equation}
\label{conv}
\Pr[t(l+k)\notin S | t(l)\notin S] \le(1-I) ^{\lfloor{l/K}\rfloor}
\end{equation}

Eqn.\ref{conv} converges to 0 for $l\rightarrow\infty $. So, the network reaches to convergence after sufficient number of iterations.

\section{Convergence Time}

The  network reaches to consensus exponentially fast for real number transfer among nodes. For averaging time $\tau$ (on each time instant/iteration step), the convergence time, $T_{\tau}$ \cite{Kashyap20071192,Boyd} is

\begin{equation}
T_{\tau}=\supinf_{t(0)} \{l:\Pr(\frac{|| t(l)-t_{ave}\mathbf{1}||_2}{||t(0)||_2})\ge \tau)\le\tau \}
\end{equation}

Let us consider, $T_1(l)$ is the random variable denoting first non-trival averaging time when $t(0)=t$. For a given graph $G$ and probability distribution of selecting $\upsilon$, lets define $\bar{T}(G)=\max [T_1(t)]$, where the maximum expectation is based over all possible initialization $t$ and $m\leq t_i\leq M, i=1,2,3,...N$. Since the minimum and maximum number of non trivial averaging are 1 and $\frac{(M-m)N}{8}$ respectively for all such initialization, the expected convergence time $T_{con}$ \cite{7068829,parvezaverage} is

\begin{equation}
\bar{T}(G) \leq \max\limits_{t:m\leq t_i\le M} E [T_{con}(t)]\leq\frac{{(M-m)^2}\times  N \times {\bar {T} (G)}}{8}
\end{equation}

\section{Simulation results}

\begin{figure*} [!htbp]
 	\centering
 	\subfigure[] { \label{fig:a}\includegraphics[width= 0.45\linewidth,height=5.5 cm]{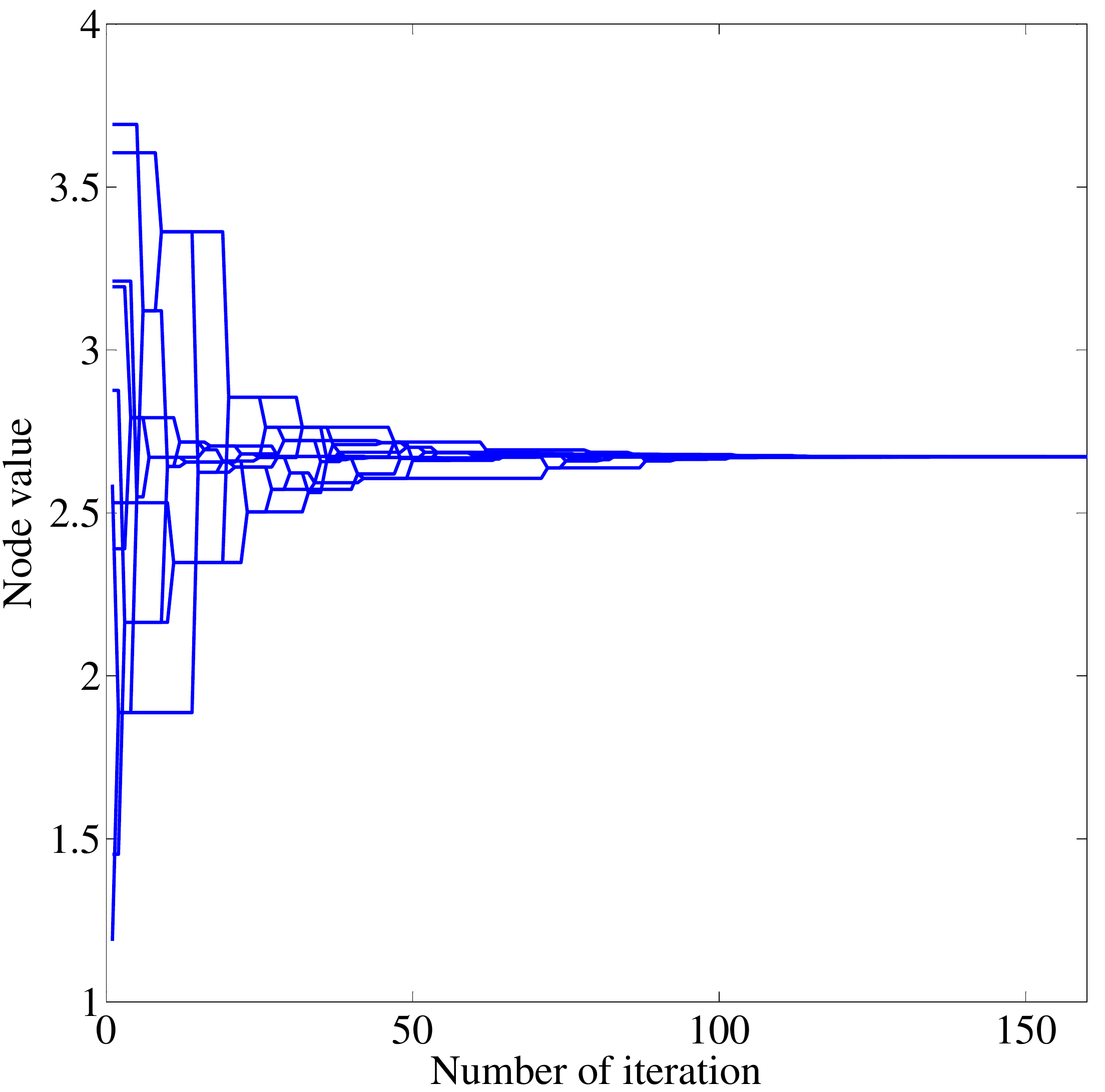}}
 	\subfigure[] {\label{fig:b}\includegraphics[width= 0.45\linewidth,height=5.5 cm]{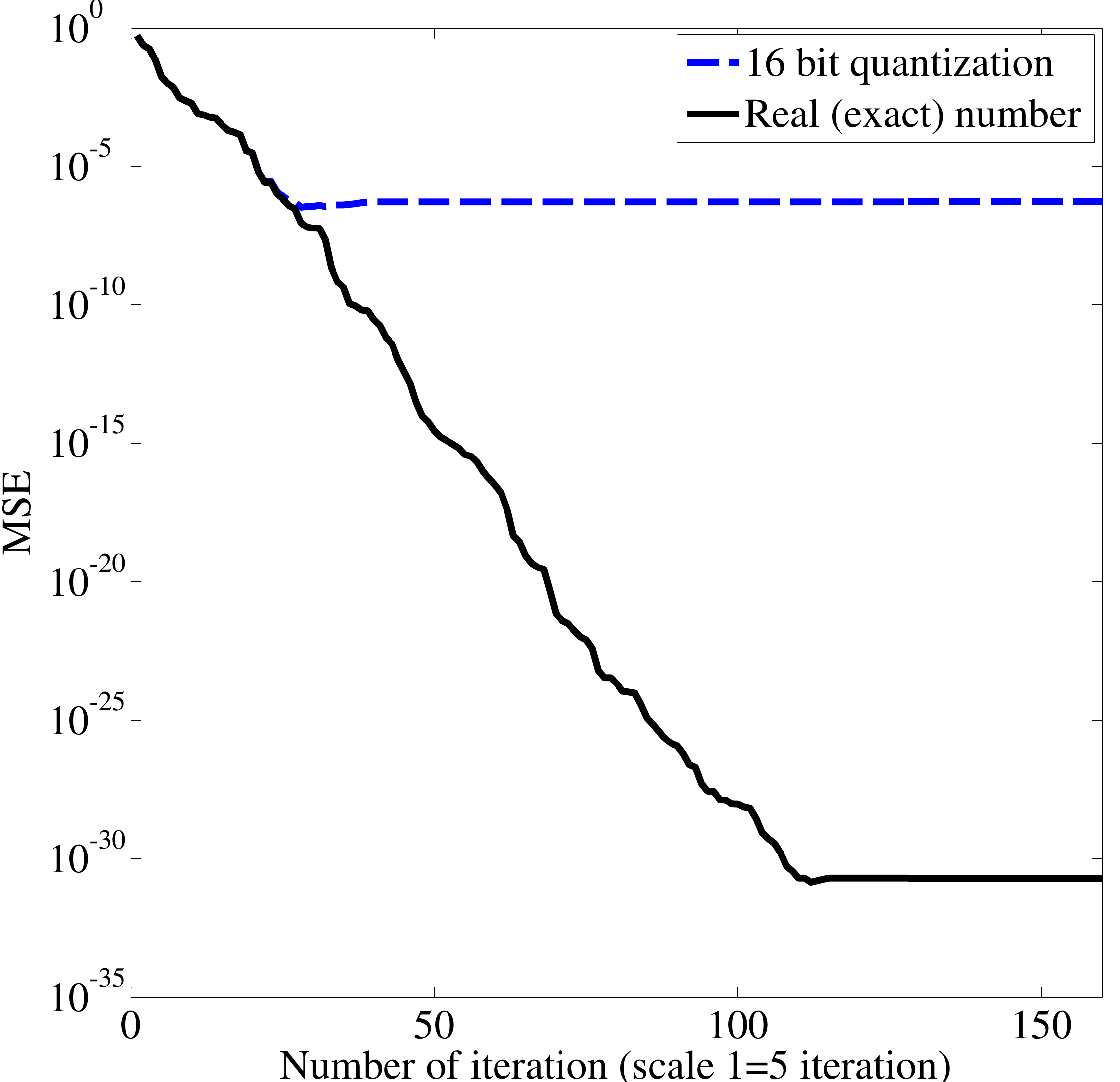}}	
 	\caption{Performance of gossip algorithm in complete graph network (a) Node value vs. Iteration number (b) MSE vs. Iteration number}
 	\label{fig:NE1}
 	\vspace*{-2mm}
 \end{figure*}

\begin{figure*} [!htbp]
 	\centering
 	\subfigure[] { \label{fig:a}\includegraphics[width= 0.45\linewidth,height=5.5 cm]{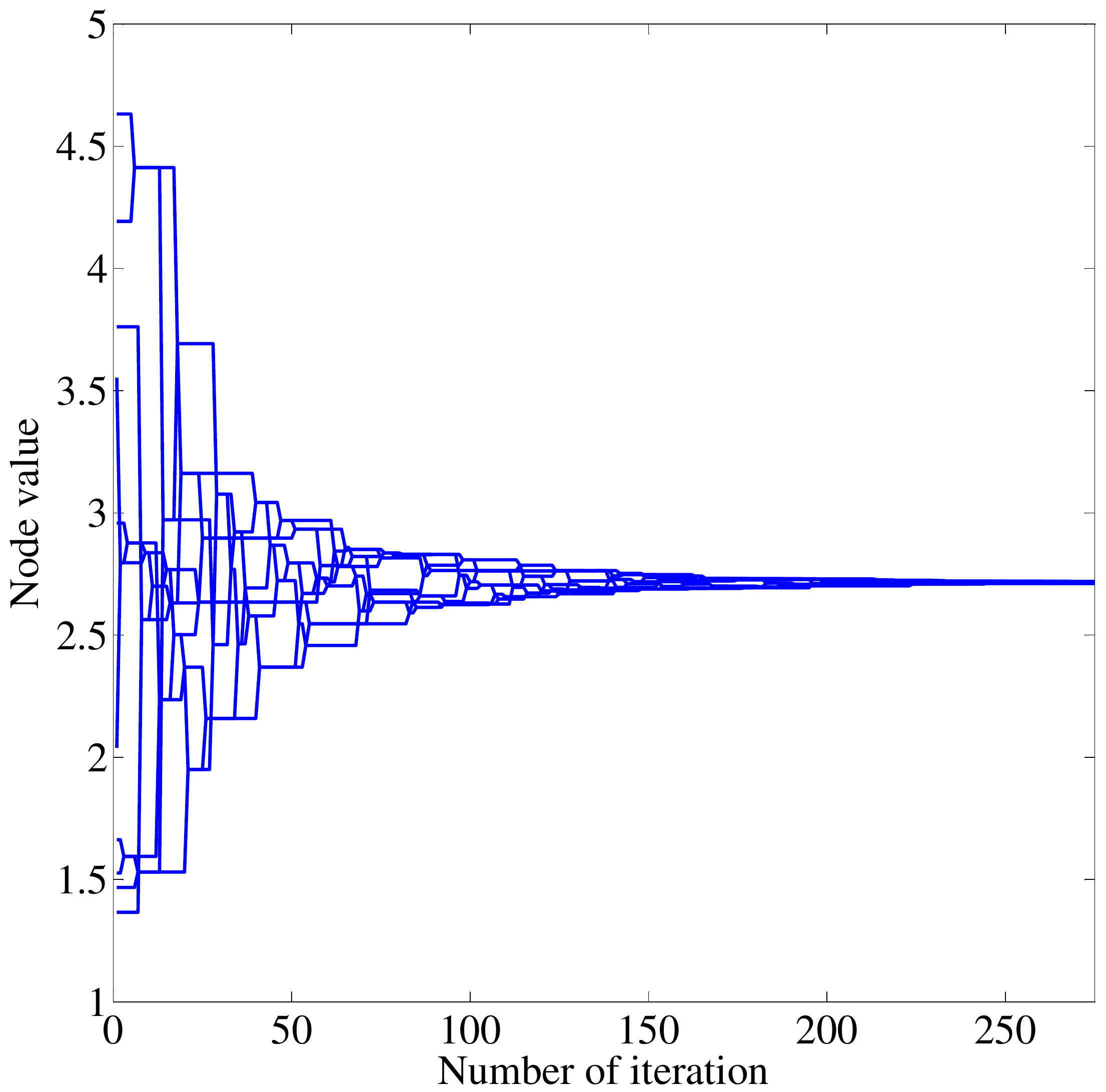}}
 	\subfigure[] {\label{fig:b}\includegraphics[width= 0.45\linewidth,height=5.5 cm]{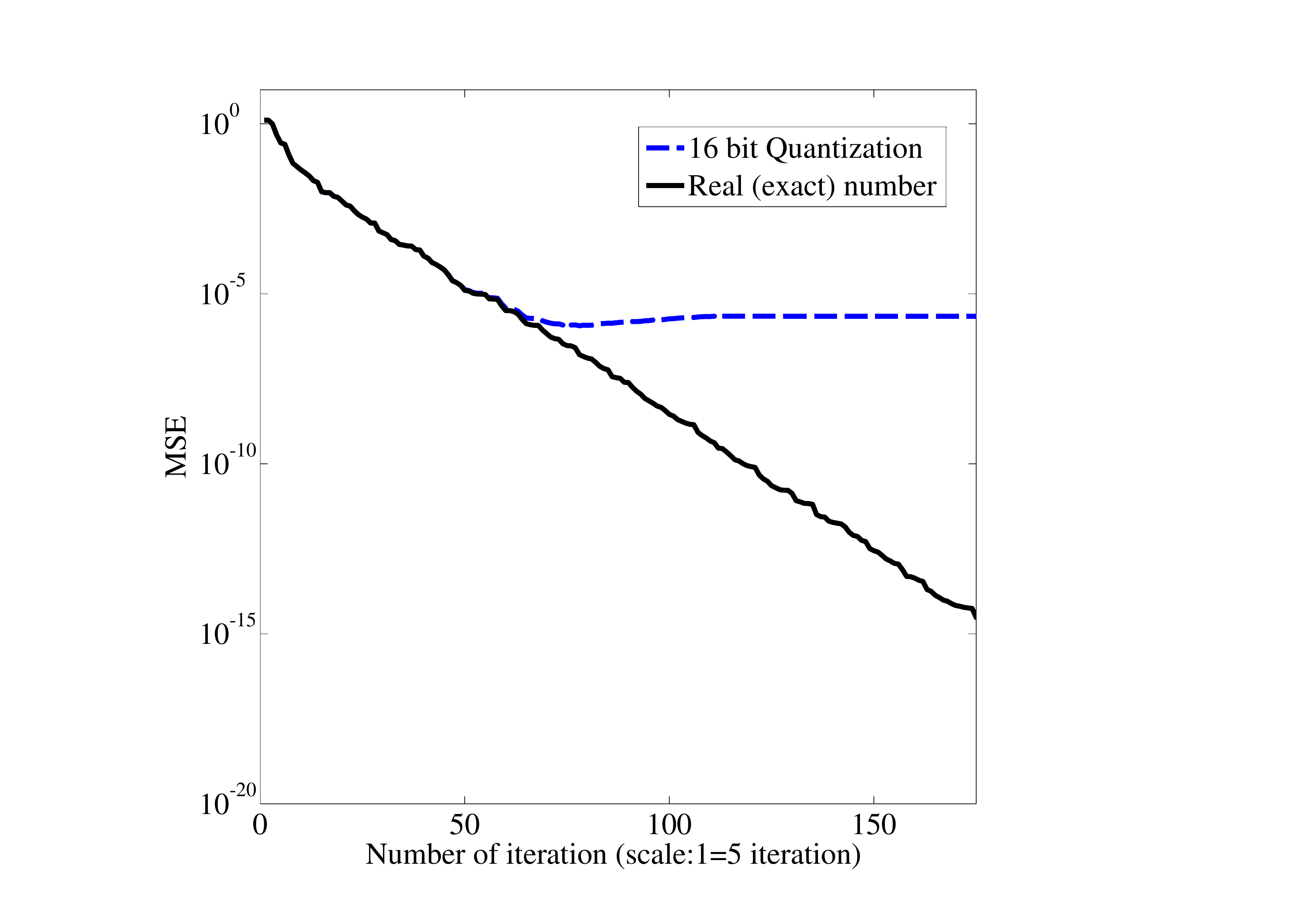}}	
 	\caption{Performance of gossip algorithm in ring graph network (a) Node value vs. Iteration number (b) MSE vs. Iteration number}
 	\label{fig:NE1}
 	\vspace*{-2mm}
 \end{figure*}

\begin{figure*} [!htbp]
 	\centering
 	\subfigure[] { \label{fig:a}\includegraphics[width= 0.45\linewidth,height=5.5 cm]{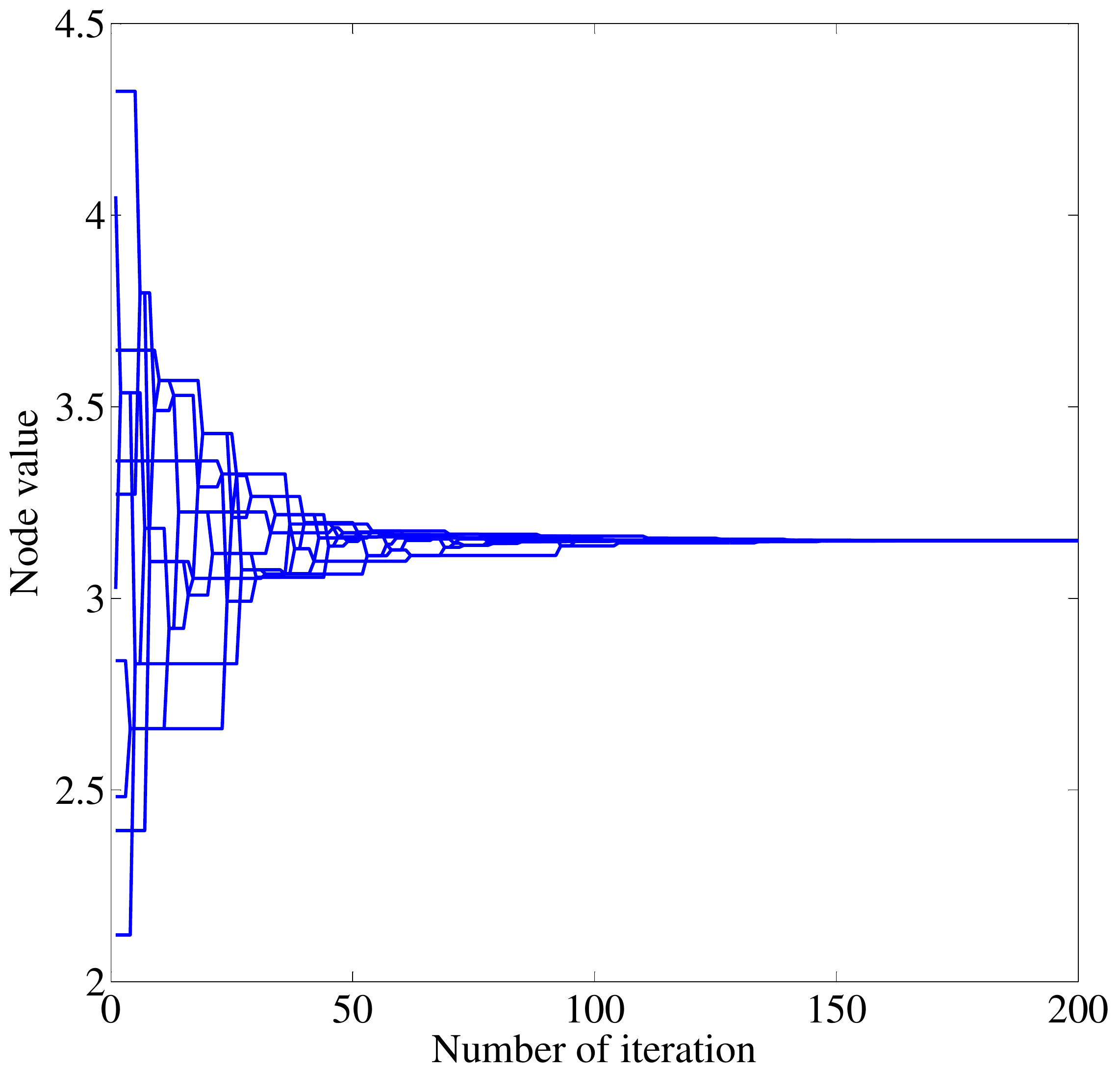}}
 	\subfigure[] {\label{fig:b}\includegraphics[width= 0.45\linewidth,height=5.5 cm]{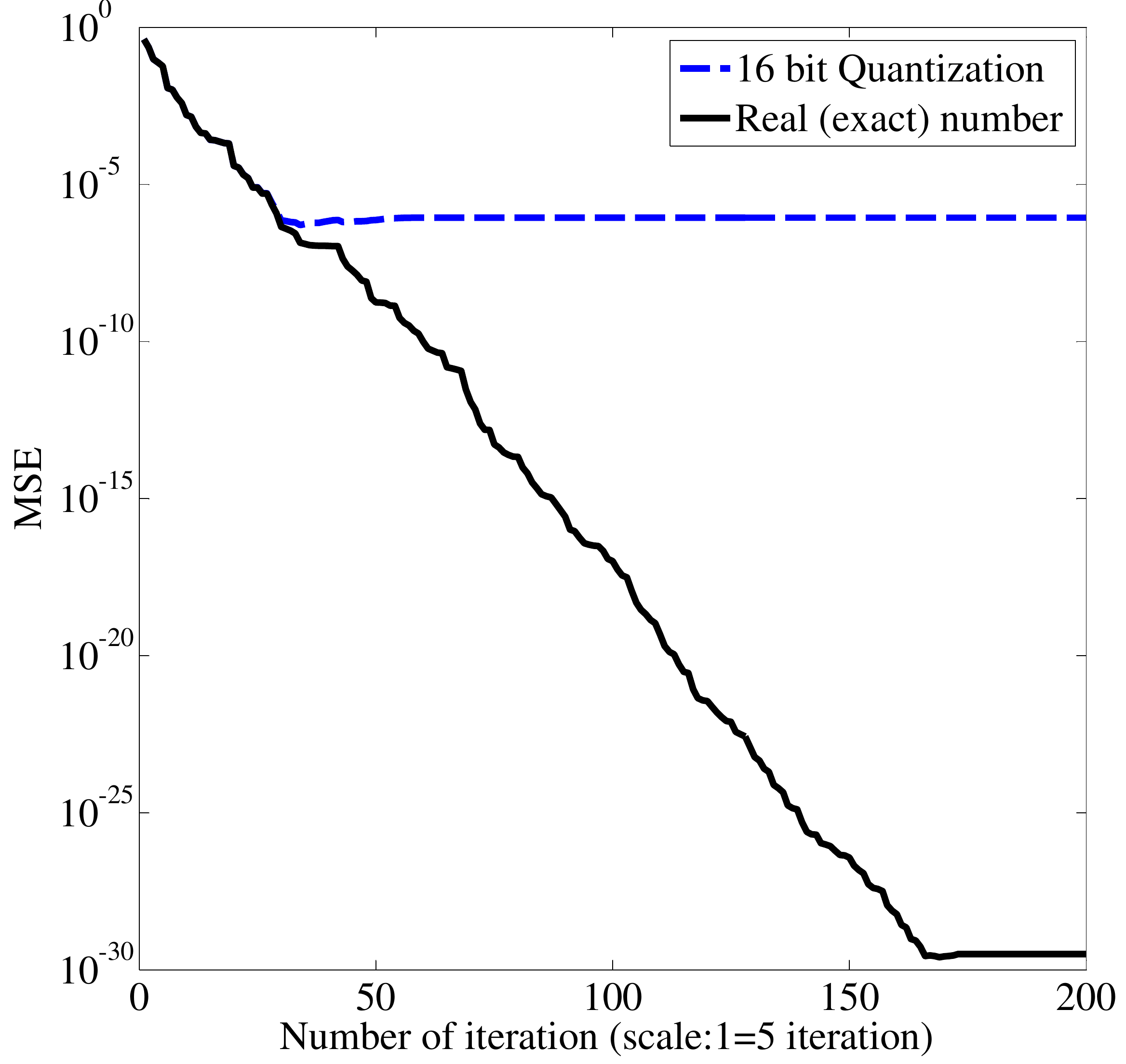}}	
 	\caption{Performance of gossip algorithm in random geometric graph network (a) Node value vs. Iteration number (b) MSE vs. Iteration number}
 	\label{fig:NE1}
 	\vspace*{-2mm}
 \end{figure*}

We simulated average consensus building through gossip algorithm for two cases: (1) real data transfer and (2) quantized data transfer. To demonstrate the performance, we consider three kind of  network topologies: complete, ring, and random geometric graph network in which nodes are placed randomly (as referred to Fig. 2). The nodes may be PMU/WAMS/smart meter or any other smart grid node. The node's time is to be synchronized. In case of wireless communication among nodes, we assume 16 bit quantized data is exchanged among nodes \cite{Thesis}.

Fig. 3 shows the performance of complete graph for 10 nodes. It is noted from Fig. 3(a) that approximately 100 iterations makes the node value to reach near consensus. On the other hand, Fig. 3(b) shows that at around 35 steps, logarithm of mean square error (MSE) becomes constant while exact node vaule transfer continues to reduce error.

The performance of gossip algorithm  is presented in Fig. 4 for star network. It is noted that the network reaches to near consensus for approximately 180 iteration steps. After 70 steps, logarithm of MSE becomes almost constant for 16 bit quantization.

The performance of gossip algorithm in random geometric graphs is shown Fig. 5. In a $1m\times 1m$ box, we generated 10 random positioned nodes, and gossip algorithm is executed between 2 nodes for distance less than 0.8m.  It is noted that the network reaches to near consensus for approximately 110 iterations and logarithm of MSE goes to constant for 40 iterations.

The relation between number of nodes and number of iterations for consensus is demonstrated in Fig. 6. It is noted that with the increase of number of nodes, the number of iterations for consensus building also increases quite linearly.

It is found that complete graph requires the least number of iterations for synchronization (consensus building). On the other hand, ring network requires the highest number of iteration steps while the iteration number for random geometric graph network depends on distance among nodes. If each iteration step takes 0.5 ms  travel time in a clustered network of 10 nodes with quantized communication, complete and ring network will provide 18~ms and 35~ms synchronization accuracy respectively. However, the number of iterations also depends on the node density and communication protocol.

\section{Conclusion}

In this study, we presented gossip based clock synchronization scheme for a  packet based smart grid network. It shows satisfactory performance for complete, ring, and random geometric graph network. Complete graph network shows the least synchronization delay (error) while ring network demonstrates the highest delay. The synchronization accuracy of random geometric graph depends on the distance among nodes. However, the synchronization accuracy depends on node density, network topology, and latency (travel time) of communication protocol.

Since in our approach, only two random nodes make average in each step, it saves bandwidth invalidating the need of  simultaneous communication among all nodes, which require huge bandwidth. In the smart grid scenario with a lot of nodes, and where GPS access might be problematic, our technique can be a potential solution for clock synchronization while maintaining quality of service (QoS).

Though we use two nodes for gossiping each time, more nodes can be used for gossiping among themselves. In the future, we look forward to investigating this along with field experiment.

\begin{figure}[!htbp]
	\centering
	\includegraphics[width= 1\linewidth,height=5.5 cm]{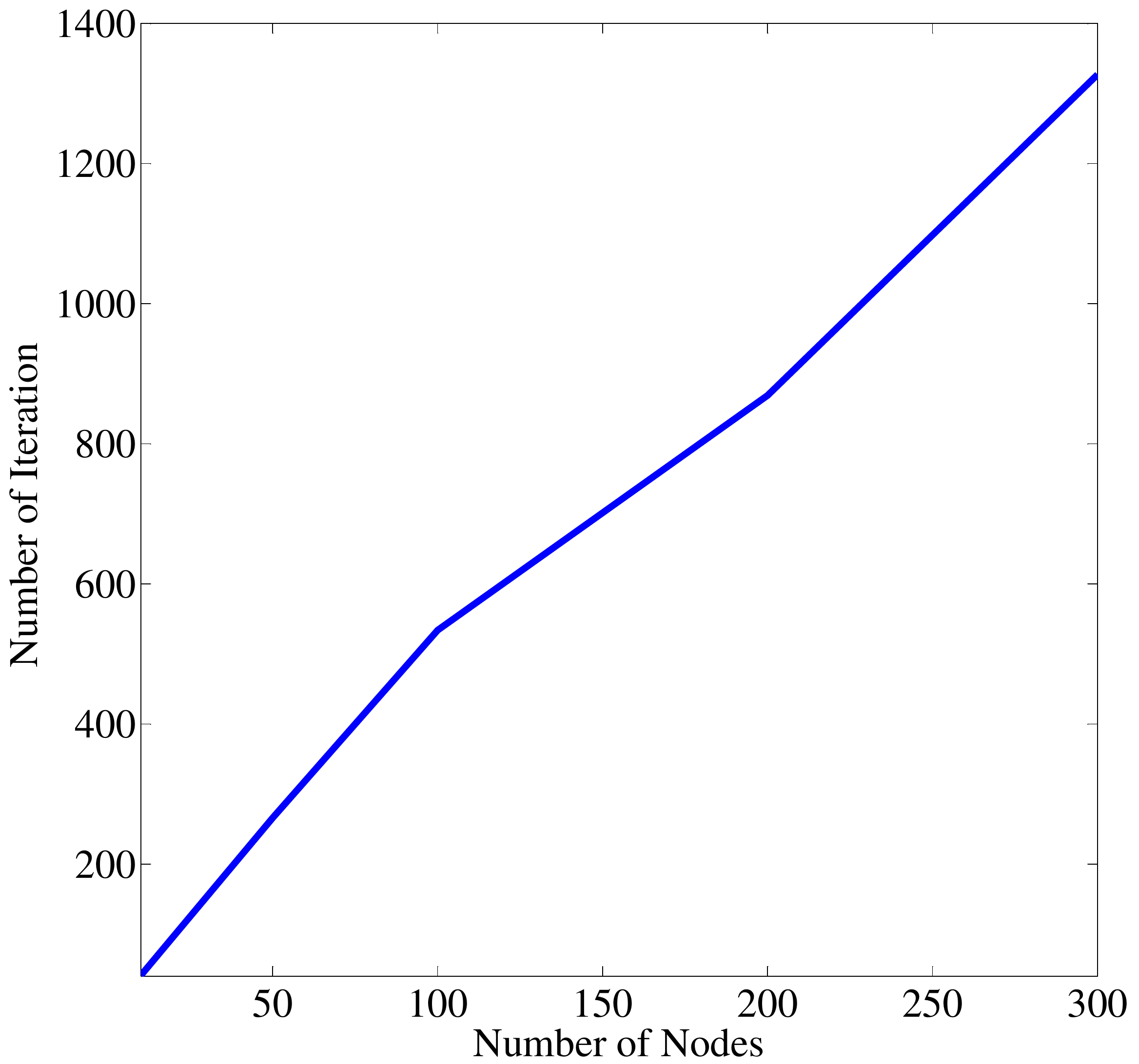}
	\caption{Convergence time/ iteration steps for different number of nodes.}
	\label{fig:roc}
\end{figure}

\bibliographystyle{IEEEtran}
\bibliography{IEEEabrv,reference}

\end{document}